\documentclass[11pt,preprint]{aastex}
\begin{document}
 
\newcommand{\ly}{Lyman}
\newcommand{\Lya}{\mbox{Ly$\alpha$}}
\newcommand{\lya}{\mbox{Ly$\alpha$}}
\newcommand{\lyb}{\mbox{Ly$\beta$}}
\newcommand{\lyg}{\mbox{Ly$\gamma$}}

\newcommand{\kms}{\mbox{km s$^{-1}$}}
\newcommand{\cmm}{\mbox{cm$^{-2}$}}
\newcommand{\cmmm}{\mbox{cm$^{-3}$}}

\newcommand{\het}{\mbox{$^3$He}}
\newcommand{\hef}{\mbox{$^4$He}}
\newcommand{\lisv}{\mbox{$^7$Li}}
\newcommand{\lisx}{\mbox{$^6$Li}}
\newcommand{\yp}{\mbox{Y$_p$}}

\newcommand{\qone}{PKS~1937--1009}
\newcommand{\qtwo}{Q1009+2956}
\newcommand{\qthree}{Q0130--4021}
\newcommand{\qfour}{HS~0105+1619}
\newcommand{\qfive}{Q1243+3047}
\newcommand{\object}{Q1243+3047}
\newcommand{\qhst}{PG~1718+4807}
\newcommand{\qpettini}{Q2206--199}
\newcommand{\qdodorico}{Q0347--3819}

\newcommand{\etal}{{\it et al.}}
\newcommand{\zabs}{\mbox{$z_{\rm abs}$}}
\newcommand{\chisq}{\mbox{$\chi^2$}}
\newcommand{\lyaf} {\lya\ forest}
\newcommand{\nhi}{\mbox{N$_{\rm H I}$}}
\newcommand{\lnhi}{\mbox{log \nhi}}
\newcommand{\ndi}{\mbox{N$_{\rm D I}$}}
\newcommand{\lndi}{\mbox{log \ndi}}
\newcommand{\bhi}{\mbox{$b_{\rm H I}$}}
\newcommand{\zdh}{\mbox{$z = 2.526$}}
\newcommand{\zdhvalue}{2.526}
\newcommand{\ETA}{\mbox{$\eta $}}
\newcommand{\ob}{\mbox{$\Omega_b$}}
\newcommand{\obh}{\mbox{$\Omega_bh^{2}$}}
\newcommand{\om}{\mbox{$\Omega_m$}}
\newcommand{\mf}{\mbox{$10^{-5}$}}


\newif\ifdraftmodep
\draftmodepfalse

\newcommand{\NOTE}[1]{\ifdraftmodep {\color {red} [{\it #1}]} \fi}
\newcommand{\DONE}[1]{\ifdraftmodep {\color {blue} [DONE: {\it #1}]} \fi}
\newcommand{\RESP}[1]{\ifdraftmodep {\color {blue} {\it #1}} \fi}

\newcommand{\NB}[1]{\ifdraftmodep {\color {red} [NB: {\it #1}]} \fi}
\newcommand{\DK}[1]{\ifdraftmodep {\color {red} [DK: {\it #1}]} \fi}
\newcommand{\DT}[1]{\ifdraftmodep {\color {red} [DT: {\it #1}]} \fi}
\newcommand{\JO}[1]{\ifdraftmodep {\color {red} [JO: {\it #1}]} \fi}
\newcommand{\NS}[1]{\ifdraftmodep {\color {red} [NS: {\it #1}]} \fi}

\newif\ifapjp
\apjpfalse

\title{Predicting QSO Continua in the \lya\ Forest}

\author{ Nao Suzuki\altaffilmark{1,2}, 
         David Tytler\altaffilmark{1,3},        
         David Kirkman\altaffilmark{1},  
         John M. O'Meara\altaffilmark{1}, 
         Dan Lubin\altaffilmark{1}  
\altaffiltext{1} {Center for Astrophysics and Space Sciences,
University of California, San Diego,
MS 0424; La Jolla; CA 92093-0424}
\altaffiltext{2} {E-mail: suzuki@ucsd.edu}
\altaffiltext{3} {E-mail: tytler@ucsd.edu}
}

\begin{abstract}
We present a method to make predictions with sets of correlated data values,
in this case QSO flux spectra. We predict the continuum in the 
Lyman-$\alpha$ forest of a QSO, from 1020 -- 1216~\AA , using
the spectrum of that QSO from 1216 -- 1600~\AA . 
We find correlations between the unabsorbed flux in these two 
wavelengths regions in the HST spectra 
of 50 QSOs. We use principal component analysis (PCA) to summarize the
variety of these spectra and we relate the weights of the principal 
components for 1020 -- 1600~\AA\ to the weights for 
1216 -- 1600 \AA , and we apply this relation to make predictions.
We test the method on the HST spectra, and we find an
average absolute flux error of 9\%, with a range 3 -- 30\%,
where individual predictions are systematically too low or too high.
We mention several ways in which the predictions might be improved.

\end{abstract}

\keywords{methods: data analysis -- methods: statistical --                     
techniques: spectroscopic -- quasars: absorption lines --         
quasars: emission lines -- intergalactic medium
}


\section{INTRODUCTION}
\label{introduction}

We would like to have an accurate and objective way to find the continuum
level in QSO spectra, because these levels are required to measure 
the amount of absorption. The uncertainty in the continuum level is often 
one of the largest 
uncertainties in the studies of intergalactic medium 
\citep[IGM, e.g.][]{croft02},
and the precise continuum shape is required 
for precision measurement of absorption lines, such as the measurement of 
D/H \citep{kirkman03,suzuki03}.

Standard methods of estimating the continua in the \lyaf\ region of QSO
spectra are frequently unsatisfactory.
For redshifts $2 < z < 4$, the standard way to find a continuum level is
to fit a smooth curve over the peaks of the flux in the \lyaf .
This method works well, giving $<2$\% errors in the continuum level
in high resolution spectra (e.g. 8 \kms\ FWHM) with high S/N
\citep[e.g. 100 per 2 \kms, ][]{kirkman03}.
However, the method fails when there are few pixels that we can 
clearly identify as unabsorbed continuum, and this lack
of continuum information is common in low S/N spectra, in 
low resolution spectra where lines blend together, and at
higher redshifts where the \lyaf\ absorbs more than a few percent at 
all wavelengths.
In fact, by redshifts $z > 6$ the complete Gunn-Peterson trough 
\citep{becker01,djorgovski01,fan01}
makes it impossible to directly measure the continuum. Instead, at high 
redshifts, especially $z > 4$, it is common to
approximate the continuum in the \lyaf\ with a power-law extrapolation 
from wavelengths larger than \lya\ 
\citep{telfer02,white03}.
But the continuum in the \lyaf\ is not a power law, because the
wings of the Ly-$\beta$-O~VI emission line and especially the \lya\ line,
extend far into the \lyaf\ and there exist weak emission lines 
especially near 1073 and 1123~\AA\
\citep{zheng97,vandenberk01,bernardi03}.

We might be able to predict the unabsorbed flux in the \lyaf\ if it is 
correlated with the unabsorbed flux at other wavelengths.
Here, the unabsorbed flux includes both continuum and emission lines, 
but not the random intervening absorption.
We measure correlations in a set of QSO spectra (hereafter the training set),
that cover both 
1020 \AA\ $ \leq \lambda \leq $ 1215 \AA\ (hereafter the blue side)
and 
1216 \AA\ $ \leq \lambda \leq $ 1600 \AA\ (hereafter the red side).
We will use the red side spectra of individual QSOs to make predictions of their
blue sides.


We use Principal Component Analysis (PCA) to summarize the information in the
QSO spectra. PCA seeks to reduce the dimensionality of large data sets
and is widely used in astronomy.
\citep{whitney83a,kanbur02,efstathiou02}.
\citet{francis92} applied PCA to the LBQS 
\citep{hewett95,hewett01}
optical spectra of QSOs to give an objective 
classification scheme, and they showed that any normalized QSO spectrum, 
$q_{i}(\lambda)$, is well represented by a reconstructed spectrum, 
$r_{i,m}(\lambda)$, that is a weighted sum of {\it m} principal components,
\begin{equation}
q_{i}(\lambda) \sim 
r_{i,m}(\lambda)=\mu(\lambda)+\sum_{j=1}^{m}c_{ij} \; \xi_{j}(\lambda),
\label{eq_reconstruction}
\end{equation}
where {\it i} refers to a QSO, ${\mu(\lambda)}$ is the mean of many QSO
spectra, ${\xi_{j}(\lambda)}$ is the {\it j} th principal component, and
$c_{ij}$ is its weight.
Instead of classifying QSO spectra, we use PCA to make predictions.

In \S 2 we describe HST spectra that we use for the training set.
We show the correlations in the QSO spectra and results of 
the PCA in \S 3. In \S 4 we show how we make predictions and we
discuss their accuracy.


\section{QSO SPECTRA AND THEIR CORRECTION}
\label{sec_2}

For the training set we use UV spectra of low redshift ($ z < 1$) QSOs 
because they have little absorption and we can clearly see their continua
levels. Here we describe these spectra, the criteria that we use to select 
them, and the corrections that we make.

We use a subset of the 334 high resolution 
{\it Hubble Space Telescope} (HST) {\it Faint Object Spectrograph} (FOS)
spectra collected and calibrated by 
\citet{bechtold02}.
This sample includes all of the high resolution 
QSO spectra from the HST QSO Absorption Line Key Project 
\citep{bahcall93,bahcall96,jannuzi98}.
The gratings chosen are G130H, G190H, and G270H, and their spectral 
resolution is R $\sim$ 1300.
\citet{bechtold02}
identified both IGM and ISM lines in a uniform 
manner and they applied Galactic extinction corrections using the 
Galactic reddening map of 
\citet*{burstein82} 
and the Milky Way reddening curve of 
\citet*{cardelli89}.

We select QSO spectra by wavelength coverage and S/N, and we remove a few 
QSOs with peculiar spectra.
We reject QSOs that do not have complete coverage from 1020 \AA\ to 1600 \AA . 
This range covers from the \lyb\ + O~VI emission line blend to the C~IV 
emission line. A larger range would help reveal the shapes of the QSO 
continua, but we would then have fewer QSOs in the sample.

We reject QSOs that did not have an average S/N $>10$ per binned pixel 
(0.5 \AA\ in the rest frame) from 1050 \AA\ to 1170 \AA .
We are interested in the intrinsic variation of the QSO spectra against
the mean spectrum. Photon noise adds variation, masks the intrinsic 
variations, and alters the primary principal components.
Before we removed the low S/N spectra, we found that some principal 
components were largely reproducing the photon noise of the spectra with 
unusually low S/N.

We remove QSOs with Broad Absorption Lines and Damped \lya\
system because we are unsure where to place their continua.
We also remove Q0219+4248 and Q0906+4305 whose emission line features
are extremely weak.
These removals make our sample not representative of all QSO spectra.

We end up using the spectra of 50 QSOs
that we list in Table 1.  
The mean redshift is 0.58,
with a standard deviation of 0.27 and a range from 0.14 to 1.04.
The average S/N is 19.5.

We will represent the spectra of all 50 QSO by fitted smooth curves
that reduce the effects of photon noise and interpolate over
the absorption lines.
To find the smooth curves, we mask the absorption lines which 
\citet{bechtold02}
identified in both the blue and red sides. 
Then, for every 50 \AA\ interval, with 20 \AA\ overlaps, we fit Chebyshev 
polynomials and we choose the order of the polynomials so that the 
reduced $\chi^{2}$ becomes close to unity.
The order is about 4 $\sim$ 6 if no strong emission line
lies in that interval.   In intervals which include strong emission lines
such as \lya\ and C~IV, the order becomes 30 $\sim$ 40. For a few QSOs
we made further adjustments by hand.

In Table 1 we give emission redshifts that we measure from the peaks of 
the \lya\ emission lines. In the rest frame the emission line peaks align, 
and the asymmetric profiles become a part of variance in the set of spectra.
If we do not use the peak of the \lya\ line, but instead we cross-correlate
with a template of known redshift, we find that we need
extra principal components to reconstruct the emission lines.
Once we obtain the redshifts, we shift the spectra to the rest frame, and 
we rebin them into 0.5 \AA\ pixels.
We then have 1161 pixels of flux data values per QSO in the range 
1020 \AA\ -- 1600 \AA . 

Since we are interested in the relative spectrum shape, we throw away
absolute flux information.  We find the average flux in 21 pixels 
around 1280~\AA , and we normalize all spectra to unit flux at these
wavelengths, far from any strong emission lines. 


\section{PRINCIPAL COMPONENT ANALYSIS OF QSO SPECTRA}

We calculate the correlation of the fluxes at different wavelengths to
see how different parts of the typical QSO spectrum are related.
In Figure 1, we see the ${1161 \times 1161}$ 
correlation matrix {\boldmath $R$} with elements 
\begin{equation}
\mbox{\boldmath $R$}(\lambda_{m},\lambda_{n})=
\frac{1}{N-1}\sum_{i=1}^{N}
\frac{\left ( q_{i}(\lambda_{m})-\mu(\lambda_{m}) \right )
      \left ( q_{i}(\lambda_{n})-\mu(\lambda_{n}) \right )}
{\sigma(\lambda_{m}) \; \sigma(\lambda_{n})},
\end{equation}
where $q_{i}(\lambda)$ is the continuum fitted and normalized 
spectrum for the {\it i} th QSO, 
$N$ is the total number of QSO spectra, 
$\sigma_{m}$ and $\sigma_{n}$ are the standard deviations of the flux 
in the {\it m}th and {\it n}th wavelength bins, 
$\lambda_{m}$ and $\lambda_{n}$ respectively.

We find moderate correlation, about 0.2 $\sim$ 0.6, between the red and blue 
continua. The correlation between the emission lines, 0.8, is much stronger, and
hence we expect that the emission lines in the red side will give 
good predictions for those in the blue side.

We can calculate the covariance matrix {\boldmath $V$} for the 50 QSOs as:
\begin{equation}
\mbox{\boldmath $V$}(\lambda_{m},\lambda_{n})=
\frac{1}{N-1}\sum_{i=1}^{N}
\left ( q_{i}(\lambda_{m})-\mu(\lambda_{m}) \right )
\left ( q_{i}(\lambda_{n})-\mu(\lambda_{n}) \right ).
\end{equation}
In Figure 2 we see the covariance is relatively small in the continuum but
large in the emission lines, meaning
that the emission lines vary a lot from QSO to QSO.  The peaks near 1073
and 1123~\AA\ probably correspond to the weak emission lines in the \lyaf\ 
that \citet{bernardi03} discuss. 

We can find the principal components by decomposing the covariance 
matrix {\boldmath $V$} into the product of the orthonormal matrix
{\boldmath $P$} which is composed of eigenvectors, and the diagonal 
matrix {\boldmath $\Lambda$} containing the eigenvalues:
\begin{equation}
\mbox{\boldmath $V$}=
\mbox{\boldmath $P$}^{-1}
\mbox{\boldmath $\Lambda$} 
\mbox{\boldmath $P$}.
\end{equation}
We call the eigenvectors, the columns of the matrix {\boldmath $P$},  
the principal components. The principal components are ordered according to
the amount of the variance in the training set that they can accommodate,
such that the first principal component is the eigenvector which has the 
largest eigenvalue.

Let us quantitatively assess how well we can reconstruct the 
QSO spectra using a certain number of the components.
We find the weight $c_{ij}$ of the {\it j}th principal component for 
QSO spectrum $q_{i}(\lambda)$ from:
\begin{equation}
c_{ij}=\int_{1020 {\rm \AA } }^{1600 {\rm \AA } }(q_{i}(\lambda)-\mu(\lambda))
\; \xi_{j}(\lambda)\; d\lambda.
\end{equation}
When we use the first $m$ components, 
we get the reconstructed spectrum $r_{i,m}$.
The ${\xi_{j}(\lambda)}$ look similar to QSO spectra, but with more structure
at the wavelengths of the emission lines.
Examples of principal components and reconstructions, which are very
similar to ours, are given by \citet{francis92}.

We now introduce the accumulated residual variance fraction $\delta E_{i,m}$:
\begin{equation}
\delta E_{i,m} =
\left .
\int_{1020 {\rm \AA } }^{1600 {\rm \AA } }
(r_{i,m}(\lambda)-q_{i}(\lambda))^2 d \lambda \right /
\int_{1020 {\rm \AA } }^{1600 {\rm \AA } }
(q_{i}(\lambda)-\mu (\lambda))^2 d \lambda .
\end{equation}
This quantity measures the square of the difference of a reconstructed
spectrum from the continuum fitted QSO spectrum, in units of the square of the
difference between that QSO and the mean. Hence
$\delta E_{i,m}$ decreases from 1 to near zero as we add components to the
reconstruction. The $m$ in $\delta E_{i,m}$ tells us that we have used the 
first $m$ components in the reconstruction $r_{i,m}(\lambda)$.
In Table 2, we list the mean
$<\delta E_{m}>  = (1/N)\sum _{i=1}^{N} \delta E_{i,m} $,
averaged over all 50 QSOs.
The first three component takes 77\% of the residual, and the first 10 components
absorb about 96\%. 
\citet[their Fig. 4]{francis92}
analyzed the LBQS set of QSO spectra 
\citep*{hewett95,hewett01}.
Using a different statistic that accounts for photon noise, different 
wavelengths, and including BAL QSOs and all absorption lines,
they found that the first three components accounted
for 75\% of the variance, and the first 10 components 95\%.
We also analyzed the LBQS spectra, kindly provided by
Paul Francis and Paul Hewett, to confirm that our implementation of the
PCA matched theirs given our wavelength range and selection criteria.
The residual variance decreases in a different way for each QSO because
the contributions of the components differ.
However, on average, the rate of reduction slows after third component 
and saturates around the 10th component. 
The components greater than about 10th look noisy and carry little information.
In the following discussion, we use up to the first 10 principal components.


\section{PREDICTING SPECTRA}

Our goal is to predict the continuum of a QSO in the \lyaf , the blue side,
using a spectrum of wavelengths larger than \lya\ emission, the red side.
In \S 4.1 we describe how we relate the blue and red side continua,
and in \S 4.2 we give a general recipe to make predictions.

\subsection{METHODS}

Unfortunately, we do not have enough QSO spectra for both a 
training set and a separate set of spectra that we can use to
test our predictions. Hence, we also use the training set for the tests.
When we make a prediction for a QSO, we use a set of principal 
components which we generated without using any spectra of that QSO.
We will call this the bootstrap method, and we will use it in most of the
remainder of this paper.
When we omit different QSOs, the first five principal components change 
by a few percent, most noticeably at the wavelengths of the emission lines.
If we do not omit the QSO of interest, the complete set of principal 
components contains all the information on each QSO, 
including the continuum fitting and photon noise. 
We found the noise features could identify a QSO, 
giving weights with unrealistically high precision.

We chose three steps to quantify the relationship between the red and
blue sides of the spectra in the training set.
\begin{itemize}
\item Step 1: Find the first $m$ principal components 
$\xi_{j}(\lambda)$ and their weights, $c_{ij}$.
We use all the blue and red side wavelengths, 1020 \AA\ to 1600 \AA , and
we use the bootstrap method to give a slightly different set of
$\xi_{j}(\lambda)$ for each QSO $i$.

\item Step 2:  Repeat step one, using only the red wavelengths
1216 \AA\ to 1600 \AA . We again keep the 
first $m$ principal components, $\zeta_{j}(\lambda)$ and 
their weights, $d_{ij}$, which are similar to those from 
step one.

\item Step 3: Solve linear equations to find a projection matrix which
relates $c_{ij}$ and $d_{ij}$.
\end{itemize}
We can write the weights in the $N \times m$ matrix form {\boldmath $C$} $= 
c_{ij}$ 
and similarly for {\boldmath $D$}.
We would like to find the $m \times m$ projection matrix 
{\boldmath $X$} $ = x_{ij}$ 
which translates weights found on the red side to the weights for
the whole spectrum:

\begin{equation}
\mbox{\boldmath $C$} = \mbox{\boldmath $D$} \cdot \mbox{\boldmath $X$}.
\end{equation}

We have $N=50$ QSOs, and we keep $m=10$ components.  
We then have more equations (N) than unknowns (m) and we wish to find the 
least-squares solution to this over-determined set of linear equations. 
The solution matrix {\boldmath $X$} can be found via 
the Singular Value Decomposition technique 
\citep{press92b}.

\subsection{MAKING A PREDICTION}
We are now ready to make a prediction of the \lyaf\ continuum of any QSO 
spectrum, that need not be in the training set, provided we have the red 
side of its spectrum.  
We obtain predictions in three steps, that are similar to those presented
above to find {\boldmath $X$}.
\begin{itemize}
\item Step A: Find the weights for the red spectrum,
\begin{equation}
b_{ij}=\int_{1216 {\rm \AA } }^{1600 {\rm \AA } }(q_{i}(\lambda)-\mu(\lambda))
\;\zeta_{j}(\lambda)\;d\lambda .
\end{equation}
The $b_{ij}$ will be like the $d_{ij}$ in step 2 above, except that
the principal components could be different. If the QSO is not
part of the training set, then the $\zeta_{j}(\lambda)$ can be derived
using the entire training set.

\item Step B: Translate the weights from the red side $b_{ij}$,
to weights for whole spectrum,
using
\begin{equation}
a_{ij}=\sum_{k=1}^{m}b_{ik} \; x_{kj}.
\end{equation}
This resembles Equation (7), except that we now know
{\boldmath $X$} and we are deriving the $a_{ij}$ that play the roles of the
$c_{ij}$ of step 1.

\item Step C: Make a predicted spectrum,
\begin{equation}
p_{i,m}(\lambda)=\mu(\lambda)+\sum_{j=1}^{m}a_{ij} \; \xi_{j}(\lambda).
\end{equation}
\end{itemize}
The predicted spectrum $p_{i,m}(\lambda)$ differs from the reconstruction
$r_{i,m}(\lambda)$ because the reconstruction uses weights
derived from the blue and red sides of the spectrum, using Equation (1), where 
as the predictions use weights derived from the red part of the spectrum alone,
using Equations (8) and (9).

We provide the two sets of principal components, $\xi_{j}(\lambda)$ and
$\zeta_{j}(\lambda)$, and the projection matrix \mbox{\boldmath $X$}, 
so the readers can make their own predictions. 


\subsection{PREDICTION ACCURACY}

In Figure 3, we show two relatively
successful and two unsuccessful predictions, $p_{i,10}(\lambda)$, 
together with the corresponding continuum fitted QSO spectra, 
$q_i(\lambda)$.  
When the prediction fails, the predicted spectrum is systematically 
either too low or too high, but with no preference.

We assess the errors on our predictions using the absolute fractional flux 
error:
\begin{equation}
\left | \delta F_{i,m} \right | =
\left .
\int_{\lambda_1}^{\lambda_2}
\left | \frac{p_{i,m}(\lambda)-q_{i}(\lambda)}{q_{i}(\lambda)} \right | 
d \lambda 
\right /
\int_{\lambda_1}^{\lambda_2} d \lambda .
\label{eq_fluxerror}
\end{equation}
For the blue side, $ \lambda_{1}=1050$ \AA\  and $ \lambda_{2}=1170$ \AA ,
avoiding the \lyb\ and \lya\ emission lines, 
and for the red side, $\lambda_{1}=1216$ \AA\ and $\lambda_{2}=1600$ \AA .
In Table 1, we list the absolute fractional flux errors for each QSO,
when we use $m=10$ components.
In Table 2, row (c) and (e), we list 
$< \left | \delta F_{m} \right | > = (1/N)\sum _{i=1}^{N}
\left | \delta F_{i,m} \right | $, 
the mean of the absolute fractional flux error from all 50 QSOs, and we 
show how this mean changes with $m$.
For comparison, in rows (b) and (d), we also list 
$< \left | \delta F_{m} \right | > $ obtained by reconstruction
when we replace $p_{i,m}(\lambda)$ in Equation (11) by 
$r_{i,m}(\lambda)$ from Equation (1). These $r_{i,m}(\lambda)$ 
use principal components derived from the blue plus red sides. We find
similar values if we use principal components from the red side alone.

The advantage of using more components is different for
the reconstruction, the red side prediction, and the blue side prediction.
For the reconstruction and the red side prediction, more components always 
improve the fit (Table 2, row (b), (c), (d)), 
although the reduction of the error is small after about the 10th component, 
and we think the remaining 3\% absolute fractional flux error is 
similar in size to the error of the continuum fitting (\S 2).

On the other hand, for the blue side prediction, adding more components
does not reduce the mean error.  
We think the reason is related to the properties of 
the first few principal components.
Only the third, fourth and fifth principal components have a significant 
slope.  If we choose appropriate weights for these components, we are likely 
to make accurate predictions, and adding more components can 
reduce the residuals (Figure 3, panel (a), (b)). 
But if we choose inappropriate
weights, there is no way to correct the slope (Figure 3, panel (c), (d)).
Hence the main error in our predictions is a systematic slope error which
makes the blue side prediction too high or too low.

Although our predictions give small errors for some QSOs, they give
huge errors for others.
With 10 components, 
the mode of the $\left | \delta F_{10} \right | $ distribution 
is 3\%, the median is 8\%, and the range is about 30\%.   
We predict the blue continua of 28 QSOs out of 50 QSOs to 
$\left | \delta F_{i} \right | \leq 10 \%$,
but we do not know which QSOs will have these small errors.
We find that many of the QSOs that give the largest errors have 
absorption or unusually low S/N on the red side of the 
\lya\ line (1216 -- 1240~\AA ,
noted in Table 1) which makes the continuum level hard to estimate. 
This region is significant because it contains much of the variance and hence
will have a large effect on the weights.
However in others cases there is little or no absorption, and the 
slope of the spectrum appears to
change as we cross the \lya\ emission line.
We do not in general see any correlation between the 
errors in the red and blue side predictions.

We introduce a third error statistic, the fractional flux error,
\begin{equation}
{\delta F_{i,m}} =
\left .
{\int_{\lambda_1}^{\lambda_2} 
\frac{p_{i,m}(\lambda)-q_{i}(\lambda)}{q_{i}(\lambda)}  d \lambda}
\right / 
\int_{\lambda_1}^{\lambda_2} d \lambda ,
\end{equation}
to help us estimate the error in the integrated flux.
This error is related to flux decrement statistic, $D_{a}$ 
\citep{oke82,bernardi03},
which has been used to calibrate numerical simulations of the IGM 
\citep{croft02}.
In Table 1 we see that 
$\delta F_{i,10} $
has a similar magnitude to 
$\left | \delta F_{i,10} \right | $, also  for the blue side,
because the predictions tend to be too low or too high across the entire blue 
side. The mean of $\delta F_{i,m} $ for all 50 QSOs,
$<\delta F_{m}> $ , is zero by definition, at each wavelength, when we use 
$m=0$, because $p_{i,0}(\lambda) = \mu (\lambda) $. When $m>0$, the statistic
$<\delta F_{m}> $ represents the bias in the
predictions, and in Table 2 we see that it remains within a few percent of 
zero for all $m \leq 10$.

In Figure 4 we show an example of the prediction of the \lyaf\ continuum
of a higher redshift QSO from the Sloan Digital Sky Survey
\citep{york00,stoughton02}
for which the continuum is unobservable
because of the large amount of absorption. To make this prediction, which
looks acceptable, we use the red side of its spectrum.


\subsection{ERRORS AND DIFFERENT METHODS}
Although the error in the levels of many of the predictions was unexpected, 
we can think of many possible explanations, including the intrinsic QSO 
spectra, calibration errors, and the method.

The intrinsic slope of the continuum may be changing at the UV wavelengths
we consider 
\citep{zheng97,telfer02}.
The UV flux from 
QSOs is black body radiation from the accretion disk around the central 
super massive black hole 
\citep{malkan83,sun89}.
If we pass over the peak of the black body continuum, the slope changes 
rapidly, and we will not have enough wavelength coverage to follow 
the change of the slope.

There are many possible errors in the calibration of the spectra,
especially the Galactic extinction correction 
\citep{fitzpatrick99}.
At our wavelengths, the extinction increases rapidly as the wavelength drops. 
If we underestimate color excess E({\it B-V}) for a $z=0$ QSO 
by 0.01 magnitudes, we decrease the flux at 1600 \AA\ by 7\%, and 
at 1020 \AA\ by 14\% . 

We have presented only one of many ways of predicting spectra.
We attempted some other less successful schemes before we arrived 
at this method.
For instance, we fitted the red side by minimizing the $\chi^{2}$ 
of $\xi_{j}(\lambda)$ against $q_{i}(\lambda)$ to get $a_{ij}$, but we 
found the blue side prediction is then unstable.
Different combinations of the principal components give similar $\chi^{2}$
on the red side, but their blue sides can have a large variety.

We also experimented with different ways of normalizing the spectra, both
near 1450~\AA\ and using the entire red side, and we obtain similar results.
We tried attempting to remove the slopes from
spectra, by fitting a straight line to the red sides, but these lines were not
adequately determined.

The predictions given here could be improved in several ways.
We should first seek improved methods, which might not involve PCA.
If we had spectra of many more QSOs we would be less subject to the
distortions and noise in individual spectra, and we could test the predictions
on spectra that were not in the training set.
Higher resolution and higher S/N spectra would help to reveal
the weak emission features, and reduce the errors in the initial continuum fit.
Extending the wavelength range on the red side may help identify the slope of 
the continuum for some QSOs, but for others we found that a reduced wavelength 
range, from 1216 -- 1400~\AA\ gave better predictions, because this restricted
range has a stronger correlation with the flux on the blue side.
We have also made predictions with the red side restricted to 
1280 -- 1600~\AA , to
avoid the large flux errors that can occur when there are absorption lines
in the \lya\ and N~V emission lines. We found that there was no significant 
change in $< \left | \delta F_{m} \right | > $.

This work was funded primarily by NASA grant NAG5-9224, and in part by
NAG5-13113, and NSF grants AST-9900842 and AST-0098731.
We are especially grateful to Paul Francis who provided the code and LBQS
spectra that he had analyzed in his 1992 paper, and who answered
many questions. Paul Hewett kindly sent the
error arrays for those spectra. Wei Zheng and Buell Januzzi kindly provided 
copies of HST QSO spectra that we used before we located the invaluable
collection of HST spectra posted to the web by Jill Bechtold.
We thank Carl Melis and Angela Chapman for their careful reading of this
manuscript.


\begin{deluxetable}{lcrrr}
\tablecaption{Statistical Quantities for 50 QSOs\tablenotemark{a} }
\tabletypesize{\tiny}
\tablewidth{0pt}
\tablehead{
\colhead{QSO} & 
\colhead{z}& 
\colhead{$|\delta F_{10}| $}&
\colhead{$|\delta F_{10}| $}&
\colhead{$ \delta F_{10}  $}\\

\colhead{} &
\colhead{}&
\colhead{Red Side}&
\colhead{Blue Side}&
\colhead{Blue Side}

}

\startdata
q0003+1553  &  0.450 &   3.3 &   5.2 &   4.1  \\
q0026+1259  &  0.145 &   5.2 &   6.7 &   5.4  \\
q0044+0303\tablenotemark{b}   &  0.623 &   4.6 &  27.3 & $-$27.1  \\
q0159$-$1147\tablenotemark{b}  &  0.669 &   2.1 &   3.1 &   2.8  \\
q0349$-$1438  &  0.615 &   2.2 &  14.7 & $-$14.5  \\
q0405$-$1219  &  0.572 &   2.3 &   4.4 &  $-$4.3  \\
q0414$-$0601  &  0.774 &   3.5 &   8.7 &   8.7  \\
q0439$-$4319  &  0.593 &   6.0 &   5.3 &   5.3  \\
q0454$-$2203\tablenotemark{b}    &  0.532 &   3.0 &  31.4 & $-$30.6  \\
q0624+6907\tablenotemark{b}    &  0.367 &   4.7 &  29.1 & $-$28.8  \\
q0637$-$7513  &  0.652 &   2.7 &  18.6 &  18.6  \\
q0923+3915  &  0.698 &   2.5 &   4.5 &   4.2  \\
q0947+3940  &  0.205 &   2.8 &  15.3 &  15.5  \\
q0953+4129\tablenotemark{b}    &  0.233 &   5.0 &  16.8 &  17.4  \\
q0954+5537\tablenotemark{b}    &  0.901 &   3.8 &   2.7 &   2.1  \\
q0959+6827  &  0.767 &   3.5 &   4.4 &  $-$2.3  \\
q1001+2910\tablenotemark{b}    &  0.328 &   5.3 &   7.9 &   7.8  \\
q1007+4147  &  0.612 &   3.4 &   8.8 &   9.1  \\
q1100+7715\tablenotemark{b}    &  0.312 &   3.9 &  31.0 &  31.0  \\
q1104+1644  &  0.630 &   3.1 &   2.7 &  $-$0.4  \\
q1115+4042\tablenotemark{b}    &  0.154 &   5.6 &  14.5 &  14.7  \\
q1137+6604\tablenotemark{b}    &  0.646 &   3.1 &   2.3 &  $-$2.3  \\
q1148+5454  &  0.970 &   3.0 &  12.3 &  12.5  \\
q1216+0655  &  0.332 &   9.4 &   4.7 &  $-$3.3  \\
q1229$-$0207\tablenotemark{b}    &  1.041 &   3.1 &   7.0 &   6.9  \\
q1248+4007\tablenotemark{b}    &  1.027 &   4.2 &   6.4 &   6.3  \\
q1252+1157\tablenotemark{b}    &  0.868 &   3.5 &   6.8 &   6.8  \\
q1259+5918\tablenotemark{b}    &  0.468 &   3.0 &  13.5 & $-$13.4  \\
q1317+2743  &  1.009 &   1.5 &  10.8 &  10.8  \\
q1320+2925\tablenotemark{b}    &  0.947 &   3.1 &   4.3 &   4.2  \\
q1322+6557  &  0.168 &   4.8 &  13.5 &  13.3  \\
q1354+1933  &  0.720 &   4.8 &   8.1 &   8.3  \\
q1402+2609\tablenotemark{b}    &  0.165 &   3.6 &  10.6 &  10.9  \\
q1424$-$1150\tablenotemark{b}    &  0.804 &   4.8 &  10.6 &  10.6  \\
q1427+4800\tablenotemark{b}    &  0.222 &   4.1 &  25.9 &  26.1  \\
q1444+4047\tablenotemark{b}    &  0.266 &   4.1 &  15.4 & $-$15.1  \\
q1538+4745\tablenotemark{b}    &  0.768 &   4.5 &  23.4 & $-$23.2  \\
q1544+4855  &  0.400 &   7.4 &  13.3 & $-$12.8  \\
q1622+2352\tablenotemark{b}    &  0.926 &   4.5 &   2.7 &   0.5  \\
q1637+5726  &  0.750 &   2.5 &   3.6 &   3.6  \\
q1821+6419\tablenotemark{b}    &  0.296 &   5.0 &  12.9 & $-$13.2  \\
q1928+7351\tablenotemark{b}    &  0.302 &   2.7 &   6.1 &  12.0  \\
q2145+0643  &  1.000 &   3.7 &   2.5 &   0.4  \\
q2201+3131\tablenotemark{b}    &  0.296 &   3.1 &   3.2 &   3.3  \\
q2243$-$1222  &  0.626 &   3.6 &   2.6 &   2.2  \\
q2251+1120\tablenotemark{b}    &  0.325 &   6.8 &  27.6 & $-$27.1  \\
q2251+1552  &  0.856 &   2.7 &   4.1 &   2.2  \\
q2340$-$0339\tablenotemark{b}    &  0.894 &   3.1 &  24.7 &  24.6  \\
q2344+0914\tablenotemark{b}    &  0.671 &   7.8 &   5.5 &   4.9  \\
q2352$-$3414  &  0.707 &   3.6 &   3.9 &   1.0  

\enddata
\tablenotetext{a}{We used 10 principal components for the error values, which we
show multiplied by 100.}
\tablenotetext{b}{The spectrum contains absorption or unusual photon noise 
in the region 1216 -- 1240~\AA\ that might have lead to an inaccurate
continuum level.}
\end{deluxetable}


\begin{deluxetable}{lrrrrrrrrrr}
\tablecaption{Mean Statistical Quantities for 50 QSOs\tablenotemark{a} } 
\tablewidth{0pt}
\tablehead{
\colhead{Component: m} &
\colhead{1}&
\colhead{2}& 
\colhead{3}&
\colhead{4}&
\colhead{5}&
\colhead{6}&
\colhead{7}&
\colhead{8}&
\colhead{9}&
\colhead{10}
}
 \startdata
\cutinhead{Blue plus Red sides:  1020 \AA\ $\leq \lambda \leq$ 1600 \AA}
(a)  Reconstruction \hspace{2.5mm} $<\delta E_{m}>$ & 50.1 & 31.5 & 22.8 & 15.9  & 11.8  
                        &  8.7 &  6.7 & 5.7 & 4.6  & 3.7 \\
\cutinhead{Red Side:   1216 \AA\ $\leq \lambda \leq$ 1600 \AA }
(b)  Reconstruction \hspace{1mm} $<|\delta F_{m}| >$  
                        & 8.9 & 7.5 & 7.7 & 5.9 & 5.1 
                        & 4.4 & 3.7 & 3.7 & 3.6 & 3.3\\
(c)  Prediction  \hspace{9.5mm} $<|\delta F_{m}|>$  
                        & 8.7 & 6.7 & 6.5 & 5.4 & 5.6 
                        & 4.4 & 4.3 & 3.9 & 4.1 & 3.1\\
\cutinhead{Blue Side:   1050 \AA\ $\leq \lambda \leq$ 1170 \AA }
(d)  Reconstruction \hspace{1mm} $<|\delta F_{m}| >$ 
                        & 10.5 & 6.9 & 5.5 & 4.1 & 4.0
                        & 3.9  & 4.1 & 3.3 & 3.7 & 3.2\\
(e)  Prediction \hspace{9.5mm} $<|\delta F_{m} | >$ 
                        & 10.4 & 9.6 & 9.7 & 9.2 & 9.6 
                        & 9.7  & 9.7 & 10.7 & 9.9 & 9.4\\
(f)  Prediction \hspace{10.5mm}  $<\delta F_{m}>$ 
                         & $-$1.4 & $-$1.0 & $-$1.2 & $-$1.0 & $-$1.2 
                         & $-$1.2 & $-$1.2 & $-$2.2 & $-$1.4 & $-$1.3\\
\enddata
\tablenotetext{a}{We have multiplied the values, which are the means for 
all 50 QSOs, by 100.}
\end{deluxetable}


\clearpage
\begin{figure}

\begin{center}
\includegraphics[scale=0.8]{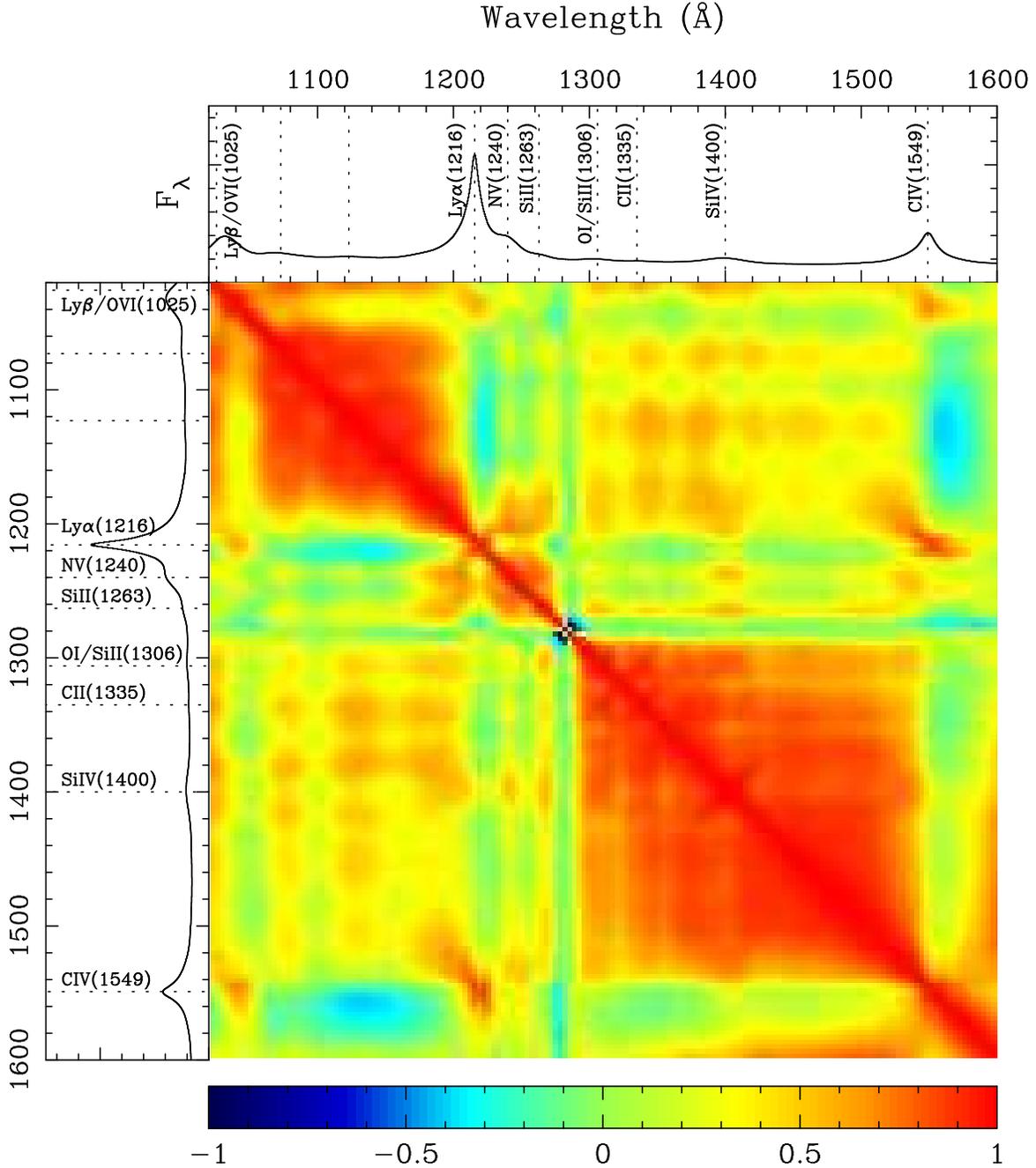}
\end{center}

\caption{
Visualization of the 1161 $\times$ 1161 correlation 
matrix (Equation (2) )
created from the continua fitted to the 50 QSO spectra,
normalized near 1280 \AA .
The numerical values, encoded in the lower bar,
depend on the normalization wavelength, and the
features near 1280 \AA\ are artifacts of the normalization.
The correlation is unity by definition on the diagonal, and the
upper and lower triangles are identical reflections about this diagonal.
Top and left panels show the wavelength range and 
the mean spectrum of the 50 QSOs.  Emission lines are shown 
at effective wavelengths from \citet{wills95}. 
We see that the correlation is strong between the emission lines, 
for example the horizontal row at the wavelength of
C~IV shows correlations 0.4 -- 0.9 at 
the wavelengths of \lyb\ and \lya .
The continuum between \lyb\ and \lya\ is moderately correlated 
(0.1 -- 0.6) 
with that between \lya\ and Si~IV, and somewhat less correlated 
(0.0 -- 0.5) with 
that from Si~IV to C~IV.
\label{fig_correlationmatrix} 
}
\end{figure}

\clearpage
\begin{figure}

\begin{center}
\includegraphics[scale=0.8]{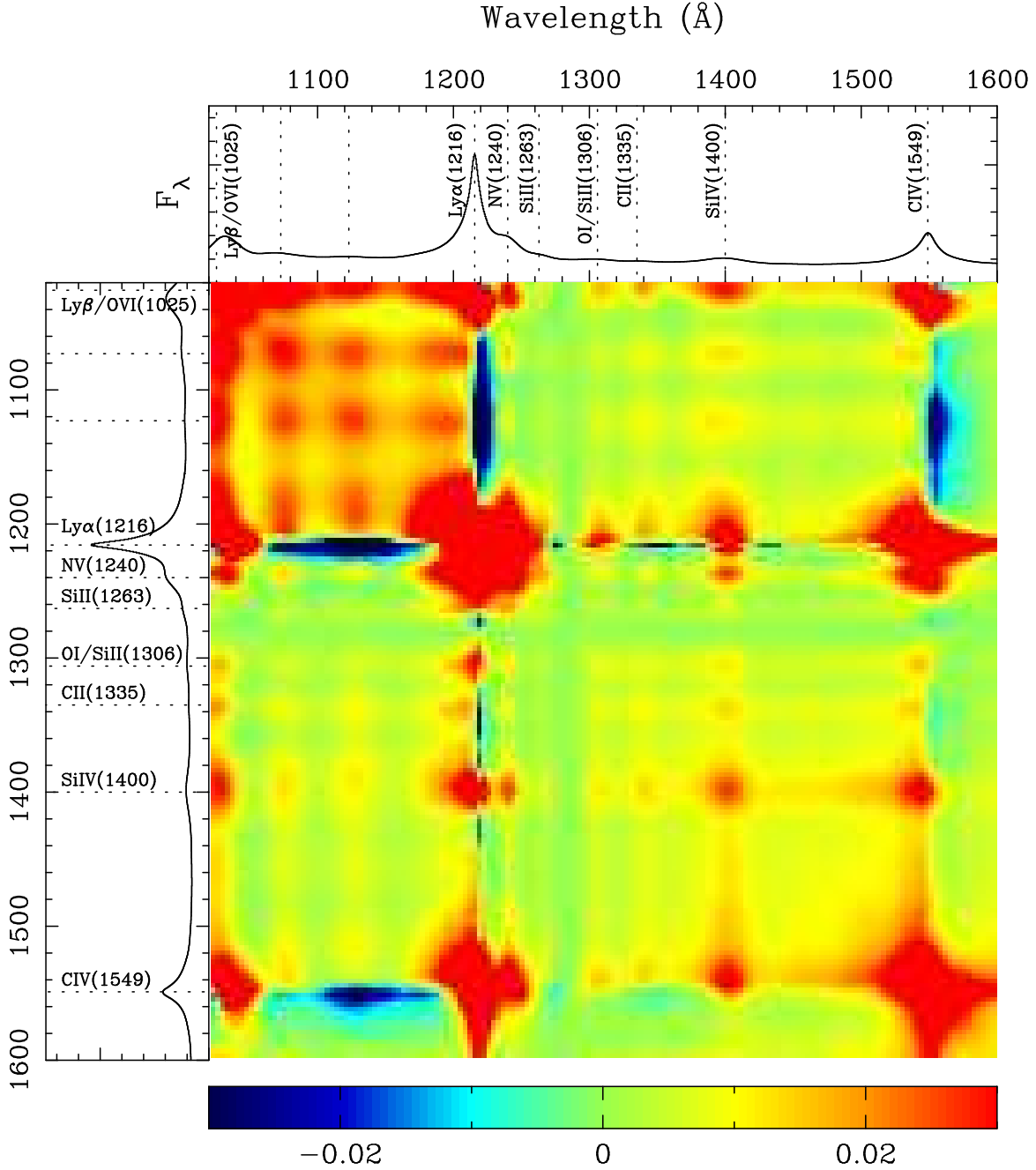}
\end{center}
\caption{
As Figure 1, except showing the covariance
matrix given by Equation (3). 
The variance is largest at the emission line wavelengths:
\lyb\ + O~VI, 1073~\AA , 1123~\AA , \lya , N~V, Si~II (1263),
O~I + Si~II (1306), C~II (1335), Si~IV + O~IV] and C~IV.
We see a grid of peaks in the variance at the intersections of these 
wavelengths.
The variance is zero at 1280~\AA\ where we normalized the spectra to the
same flux.
The variance is larger between \lyb\ and \lya\ than on the red side, 
perhaps because of errors in our continuum fits in the \lyaf\ or 
because of intrinsic variations.
We obtain the principal components by decomposing 
this covariance matrix into eigenvectors and eigenvalues.
\label{fig_covariancematrix}  }
\end{figure}

\clearpage
\begin{figure}
\begin{center}
\includegraphics[scale=0.6]{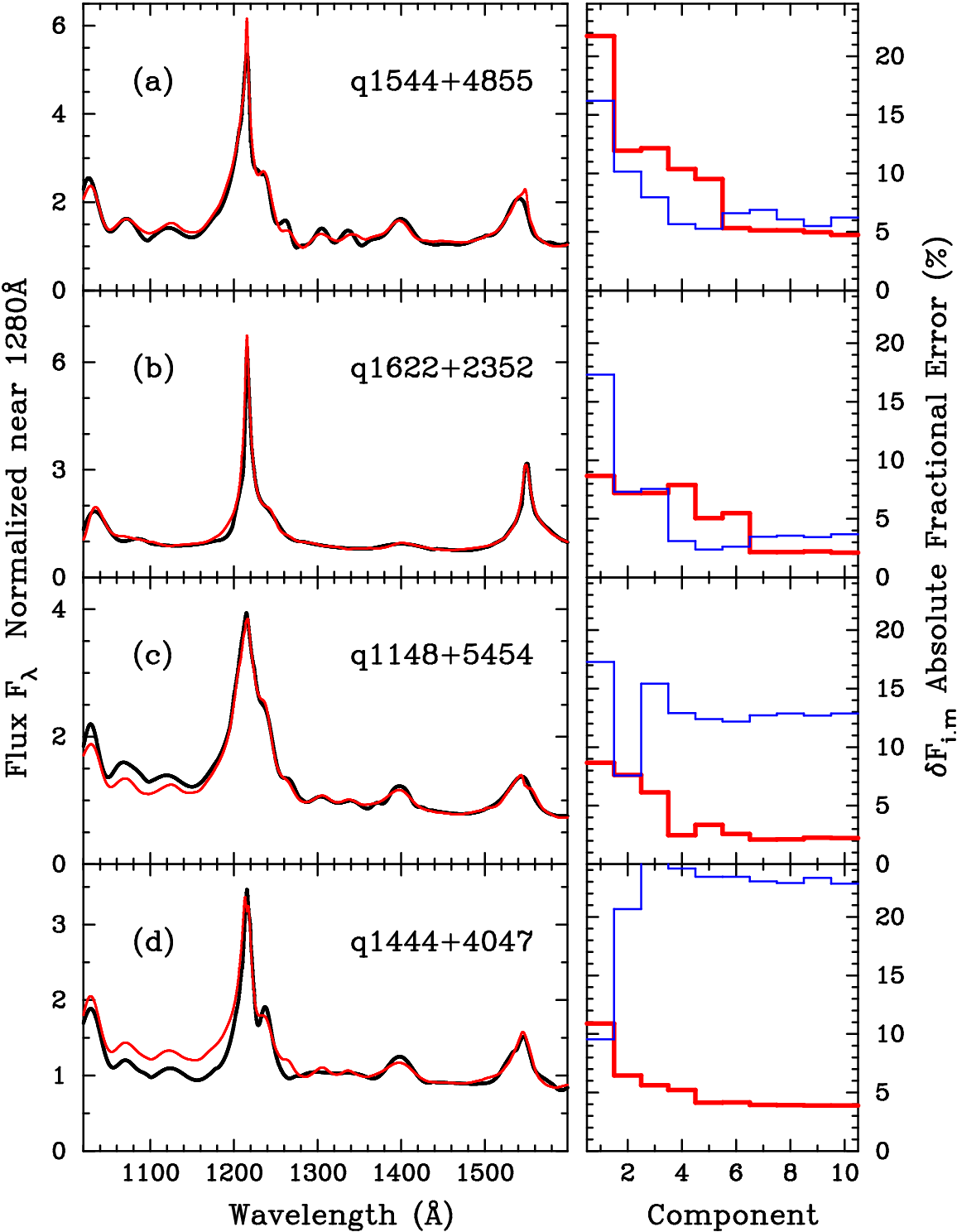}
\end{center}
\caption{
Four examples of our predicted spectra and their accuracy.
The top two gave relatively small errors, and the bottom two 
relatively large. 
Note that the vertical scales are not all the same.
The panels on the left show 
in thick lines the original continuum fitted spectra, $q_{i}(\lambda)$.
The thin lines show the spectra predicted using 10 components 
($p_{i,10}(\lambda)$, Equation (10)).
We do not show any reconstructions $r_{i,j}(\lambda)$, to 
either the red sides or the whole spectra.
The panels on the right show the corresponding absolute fractional
flux error (Equation (11)) averaged over the part of the blue 
(1050 -- 1170~\AA , thin lines) and the whole of the red (thick) 
sides for each QSO.
For both the blue and red we show the predictions (Equation (10)) and not
reconstructions, which are usually better. 
For the top two QSOs the errors decrease when we use more components,
reaching 2 -- 6\% on both the red and blue sides with 10 components.
For the bottom two QSOs, the predictions reach 3 -- 4\% on the red side,
because we use information from those parts of the spectrum, but they
are too low or too high by 13 -- 23 \% on the blue side, primarily
because the level is wrong, even though the emission line shapes seen
excellent. 
These four QSOs represent the variety found in the sample of 50 
QSOs. We were unable to predict which QSOs would behave like the 
top or bottom two.
\label{fig_demo}  }
\end{figure}

\clearpage
\begin{figure}
\begin{center}
\includegraphics[angle=270,scale=0.6]{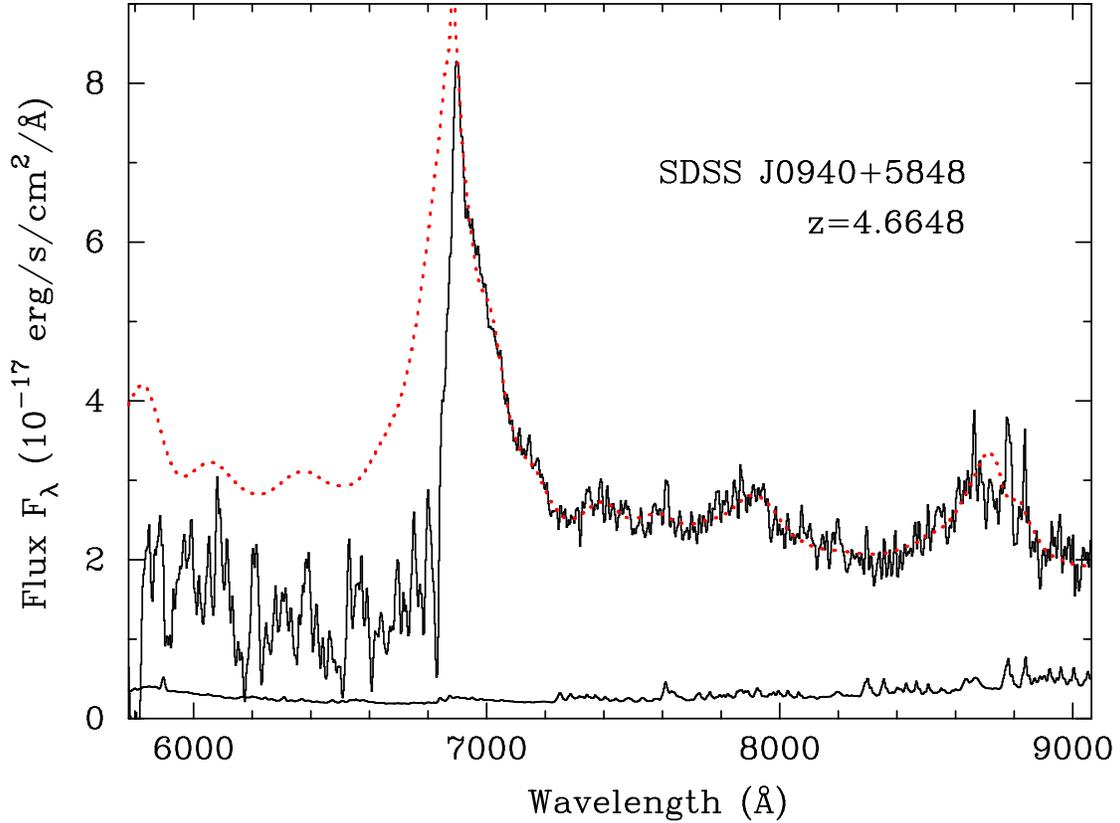}
\end{center}
\caption{
An example of a continuum predicted (dotted line) 
using Equations (8) -- (10) for a high redshift QSO, 
SDSS J0940+5848 (solid lines). 
The only information that
we use on this QSO is the red side of this spectrum, from 1216 -- 1600~\AA .
The \lyaf\ forest absorbs strongly and no pixels in this low resolution
spectrum reach the predicted continuum level. We suspect the predicted
blue continuum has the correct shape, including the two emission lines
between \lyb\ and \lya , but we know that the overall level could be in error
by 5 -- 30\%, as we saw in Figure 3.
\label{fig_sdss}  }
\end{figure}

\end{document}